\documentclass{eas}
\usepackage{graphicx}
\begin{document}

\title{Review on non-directional direct dark matter searches} 
\author{B. Censier}\address{Institut de Physique Nucleaire de Lyon, Universite de Lyon, Universite Lyon 1, CNRS/IN2P3,
4 rue E. Fermi 69622 Villeurbanne cedex, FRANCE}
\begin{abstract}
An overview of non-directional direct detection methods is given. The currently leading experiments for spin independent WIMPs interactions are using simultaneous measurement of two quantities for event-by-event background discrimination in cryogenic bolometers and noble gas like xenon. Besides these, several interesting techniques have been developped, each having a specific advantage concerning e.g energy threshold lowering or strong immunity to ionizing radiations background. Technologies used and most recent results about spin-dependent and spin-independent cases are presented.
\end{abstract}
\maketitle
\section{Introduction}
We will present here the most salient facts and results concerning the (non-directional) hunt for Weakly Interactive Massive Particles (WIMPs), most often presented within the framework of Minimal SupeSymetric Models (MSSM) which offers a well suited candidate (the neutralino as a heavy, stable particle), and more pragmatically, a benchmark for experiments sensitivity comparison. The candidate to be detected has the following basic properties: it should be neutral, quite heavy (O(100GeV)), and interacts very weakly with "ordinary" matter.
From an experimental point of view, this in turns set the main constraints for an experiment design, namely a ~keV to 10 keV energy threshold, and a small background counting rate (less than 0.01 event/kg/day from latest experimental constraints). These two constraints call for some massive (O(1kg) or more) and sensitive detectors, the main difficulty already arising being the opposition between mass and sensitivity for most of the detection methods. Given the expected counting rates, one has to expose detectors to the predicted WIMPs background during long period (O(1 year)), which requires good long term stability by regular calibrations and/or a precise control over several environmental parameters. Among these environmental parameters, one of the most important is of course the particles background at these energies, from ambient radioactivity and from the influence of cosmic rays. The latter problem is partly adressed by installing experiment in underground laboratory and using passive shielding, the former by carefully controlling the radiopurity of the experiment ans its vicinity. One important features with respect to neutral particles search is the ability of adding an active rejection method for ionizing particles in addition to the passive shielding. After presenting the main facts about background discrimination and possible observational signatures, we will detail major results concerning spin-independent coupling between WIMPs and target nuclei, followed by major results concerning the spin-dependent case. We finally sum up some observations and results about the recent "low mass" (O(1GeV)) hints that some experiments have observed.

\section{Dark matter direct detection}
\subsection{Detection channels and background discrimination}
\begin{figure}
\begin{center}
\begin{tabular}{cc}
\includegraphics[%
  width=0.5\linewidth,
  keepaspectratio]{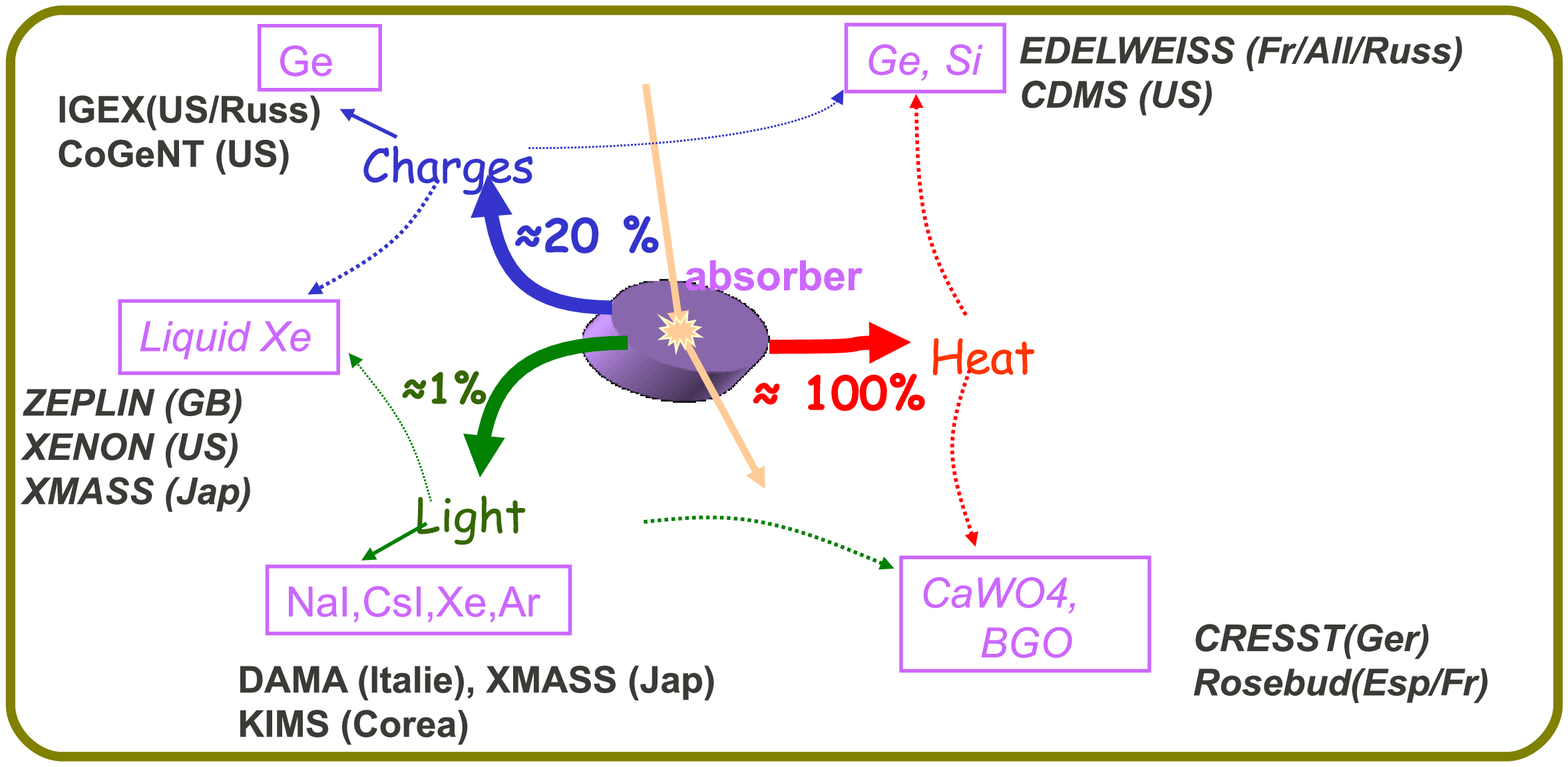}&
\includegraphics[%
  width=0.5\linewidth,
  keepaspectratio]{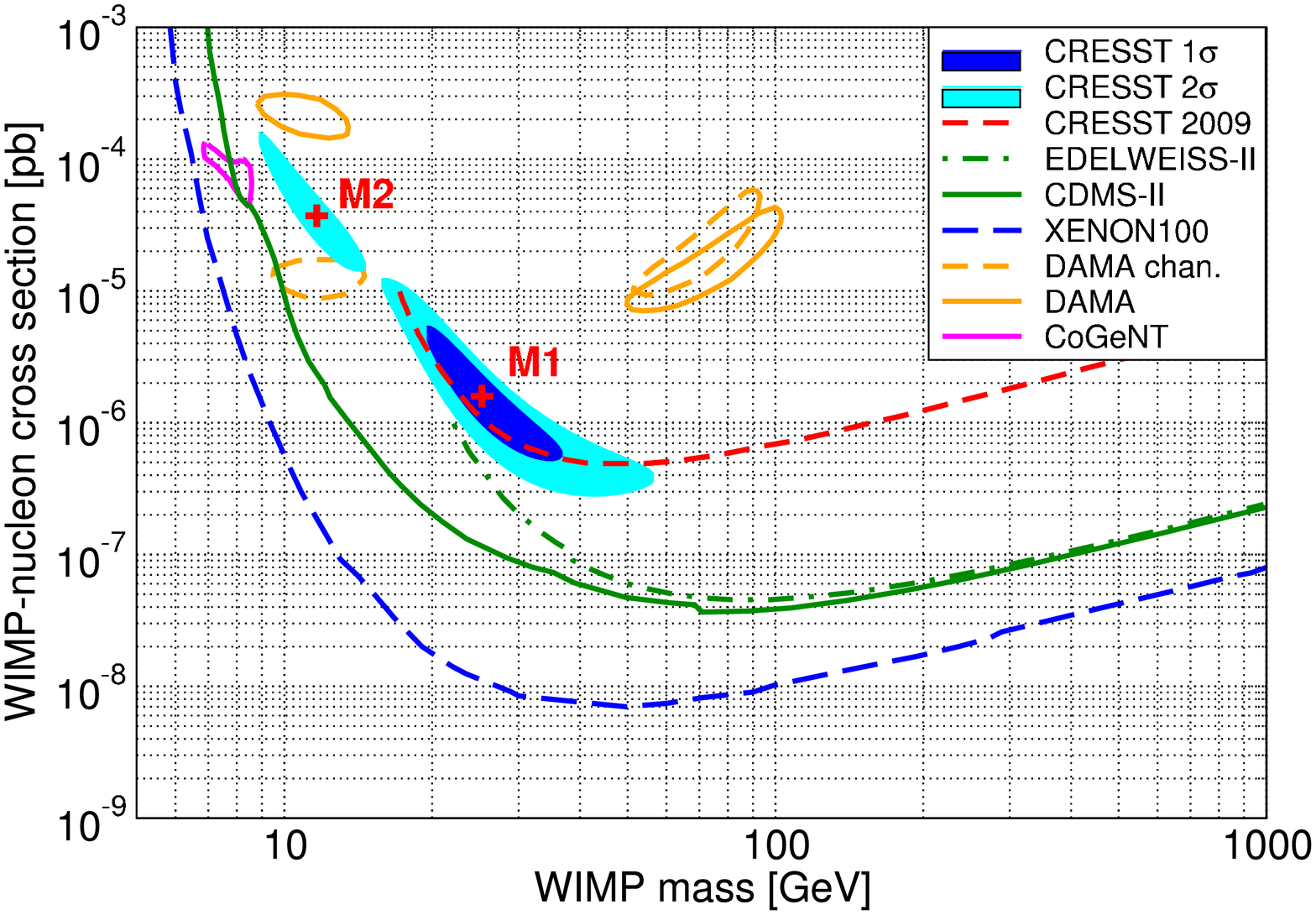}
\end{tabular}
\end{center}
\caption{Left panel: Possible detection channels in several absorbers for an initial energy deposit by an incoming particle. The experiments listed in the picture use one channel (ex: DAMA, IGEX, CoGeNT), or two simultaneous channels measurement (ex: EDELWEISS, XENON, CRESST, CDMS). The latter feature allows for ionizing background rejection. Not shown here, there also exist interesting methods based on the use of metastable materials (ex: bubble chamber for COUPP, superheated droplets for PICASSO). Right panel: latest constraints for spin-independent interactions in the MSSM phase space \cite{CRESST}}
\label{detection_channels}
\end{figure}

As pictured in figure \ref{detection_channels}, the energy deposited by an incoming particles inside an absorber may be measured under three main forms: heat, scintillation or ionisation. The pertinency of using one or several of these three channel obviously depend on the absorber type. Bolometry allows heat measurement, this channel giving the total thermalized energy of all channels when observed on a sufficiently long time-scale (i.e the time-scale for thermalization). This channel will thus be measuring nearly all the energy deposited. As an exception to that, if scintillation processes take place in a transparent absorber, there is about 1$\%$ or so of the deposited energy that will not be thermalized and will be lost for the heat channel. The scintillation photons may then be detected by a separate photodetector. Scintillation is closely linked to excitation of electronic states and to the following photons emission. Its intensity may be dependent on the spatial density of the deposited energy, thus on stopping power and on energy on the incident particle. Finally, part of the energy deposited may be absorbed in atom ionisations (typically 10$\%$ or so). In a semiconductor absorber, an energy greater than the effective energy gap beetwen valence and conduction bands will lead to the generation of electron-hole pairs that can be collected by biased electrodes and measured through associated charge read-out.
For the first generation of direct dark matter detection, e.g. IGEX, the theoretical expectations on WIMPs cross-section was sufficiently loose so that a single channel measurement, without any identification of ionizing background, was sufficient to improve existing constraints.
A second generation, e.g. CDMS, CRESST, EDELWEISS or XENON, began to use an event-by-event discrimination based on the simultaneous measurement of two channels. The ratio of energies deposited in each channel can then serve as a powerful discriminator between ionizing (background from so-called electron recoils) and non-ionizing (WIMPs candidates from so-called nuclear recoils) particles. CDMS and EDELWEISS both use simultaneous measurement of heat and ionisation in low-temperature Ge absorbers. CRESST uses simultaneous measurement of heat and scintillation in low temperature C$_a$WO$_4$, while XENON uses ionisation and scintillation detection in two phases liquid/gas xenon tank.
On the other hand, some experiments (e.g. DAMA or CoGeNT) took different design choices. While they use single channel measurement, preventing them to get some precious informations about ionizing background, they are able to build simpler setups, which in turn may allow to have bigger mass and/or better features like improved energy resolution by loosening technical constraints.
Depending on these design choices, an experiment will be able to sign a candidate detection from one or several signatures.

\subsection{(non-directional) Signatures}
Let us first talk about experiments with ionizing background rejection.
In such an experiment, the WIMPs candidates are identified by requiring them to be of the nuclear recoil type. In addition to this selection, several additional cuts are made to select the candidate events. One of this cut is present in lots of experiments: this is the one allowing to define a so-called fiducial volume inside the absorbers. One has indeed generally much interests in rejecting events happening in the outer parts of the detector. First it allows to reject most of the low energy, low penetrating background events from external sources like ambiant radioactivity. Second it allows to overcome some physical limitations of the measurement method near the surface of the detector. 
Once the candidates are selected, one could think of using their energy spectrum shape to confirm a WIMP detection. This is practically rather difficult, the expected energy spectrum in preferred MSSM models having a quite universal exponential shape towards low energy. One should nevertheless note that in the case of an inelastic interaction, characterized by a given energy threshold under which no energy transfer can take place between an incident WIMP and a target nucleus, the candidates energy spectrum could exhibit a distinctive feature, namely a rise toward low energy but with a low energy frontier below which there is no events.
 Finally, an interesting option for a clear confirmation of detection would be the comparison between several absorbers. This could allow characterizing more precisely the type of coupling the candidates have with target absorbers. The existence of several experiments using different methods should already fulfill this requirement. For the expected 1 ton phases of current experiments though, the use of several different absorbers inside a single experiment and/or site would certainly be much useful.
 
 Concerning experiments without background discrimination, the signatures can only be found on some data gathering candidates as well as background events. A first possible signature would be an excess rate in the low energy spectrum with respect to a given background model. Given the very low background rate involved, this model is hard to obtained from simulations or observations. It thus also implies some hypothesis on the background.
 A seconde possible signature is the detection of an annualy modulated event rate in a given energy bin, resulting from the motion of earth around sun and of sun around galactic center, which are modulating the relative velocity with respect to the dark matter halo and thus the measured recoil energy. 
 It is a rather robust signature with respect to theoretical models, the effect being predicted on basic cinematics grounds. Nevertheless, it may not be so robust on the experimental point of view, several environmental parameters being annulay modulated. This signature is thus adressing the delicate problem of long term systematic errors.

\subsection{Spin-independent coupling}


Latest results on constraining spin-independent coupling are showed in figure \ref{spin_independent_exclusions}. The stronger constraints on this case are now set by the XENON experiment \cite{XENON}. Noble gas experiments (Ar or Xe) are more and more present in the direct detection field. One can differentiate single phase and douple phase ones. 

In the former case, the only detection channel is the scintillation of a liquid absorber surrounded by photomultipliers (XMASS, CLEAN/DEAP). No additional measurement channel can be used for ionising background discrimination. Nevertheless, there are some possibilities of using pulse shape discrimination methods. 

In the double phase case (XENON, LUX, ZEPLIN), free charges generated by an energy deposit inside the liquid phase can be collected by an electric field through this liquid, and then be multiplied by impact ionisation in a strong electric field (~kV/cm) when reaching an additional gaseous phase above the liquid one. 
In the XENON 100 experiment, the absorber is a 170 kg liquid xenon tank (65 kg fiducial) fitted with photomultipliers: 80 are immersed at the bottom of the absorber (S1 channel), 98 are above the gaseous phase (S2 channel). S1 allows th detection of scintillation caused by energy deposit inside the liquid Xe absorber, while S2 detects scintillation arising from the electron multiplication process in the gaseous phase. The S1 signal is a prompt one, setting the start time of an event, while the S2 signal is a delayed one due to the drift time of charges (proportional scintillation). S1 allows to getthe radial position of each event by triangulation with a precision of about 3 mm, and the delay between S1 and S2 gives the z position with a ~2 mm precision.  Due to their different ionization densities, electron recoils and nuclear recoils have a different S2/S1 ratio, which is used as ionizing background discrimination parameter. The latest XENON results correpsond to a 1471 kg.days exposure (45kg fiducial mass for 100.1 live days).

Besides XENON, the best current constraints are set by two experiments using heat-and-ionisation ultra-pure germanium bolometer, namely CDMS and EDELWEISS. The both need low temperature operations at the 10 mK level due to thermal noise reduction constraints on the heat channel. CDMS uses low impedance transition edge sensors (TESs) for this heat channel, while EDELWEISS uses high impedance Ge-NTD (neutron transmutation doped) thermistors. This imply some fundamental differences in the heat read-out electronics. Concerning the ionisation channel, at such low temperatures the semiconducting germanium absorber is already fully depleted even without any electric field applied. this allows to apply electric field as low as 1 V/cm or less for charge collection through the germanium, up to collection electrodes. The intensity of this collection field is anyway constrained by the Luke-Neganov effect, for which energy dissipation of the free carriers accelerated by the collection field contribute to the total heat measured on the heat channel. Since the discrimination method is based on the ratio of ionisation over heat energy, a too strong collection field would imply a too important heating from drifting charges: both channels would be too much correlated for the discrimination method to be efficient. For an electron recoil, the whole incoming energy is deposited in the ionisation channel and then measured on the heat channel after thermalisation has been completed: it thus gives an ionisation over heat close to 1. For a nuclear recoil, the whole deposited energy first goes into a target nucleus displacement inside the regular Ge cristal. In this primary phase of energy dissipation, this target nucleus transitorily acts as an ion penetrating the absorber with the energy received from the incident particle. Part of this energy will thus be dissipated by cristal collective modes excitation (phonons emission), while another part will be dissipated by ionisation of nearby atoms. In this case, the ionisation over heat ratio is measured at about 0.3.
The global picture for both experiments is thus an ionisation signal being dependent on the interaction type, and a heat signal giving an almost absolute energy scale regardless of the interaction. This technique allows to reach energy resolutions of the order of 1 keV for 400 to 800g detectors. The trigger signal is taken based on the heat channel, due to the reduced ionisation energy for nuclear recoils.
The definition of a fiducial volume is roughly the same in both experiments: it is based on the ionisation channel by segmenting electrodes and allowing the rejection of events happening in the outer part of a cylindrical detector (guard ring electrodes). Nevertheless, the fiducialisation is not complete with that technique, since it lacks the rejection of near electrodes regions with respect to the z coordinate. The first phases of these experiments showed this was one of the biggest limitations, near electrodes events showing a charge collection deficit. Each of the two experiments now use an additional cut for rejection of these events.
In CDMS, the cut is based on heat channel timing (rise time and delay with ionisation channel). This can be done because the TESs are sensitive to so-called out-of-equilibrium phonons that are emitted before thermalisation is reached.
In EDELWEISS, it is based on the ionisation channel, by continuation of the electrode segmenting solution. Top and bottom planar electrodes on the cylindrical absorber are replaced by concentric ring electrodes, alternatively set to opposite sign voltages. One over two electrodes is part of the "collection electrode" while the others are the "veto electrode". For an event inside the bulk, the electric field lines are almost vertical, and collection is made between the top and bottom collection electrodes: there is a signal on the collection channel, but not on the veto one. For a near-surface event ($\approx$1 mm from the top or bottom surface), the field line topology is rather horizontal by joining two adjacent collection and veto rings: there is a signal on the veto channel. Data cuts based on this channel thus allow to reject near-surface events.
EDELWEISS, CDMS and XENON, all have detected some candidates. The statistics is nevertheless insufficient for concluding to the detection of a new background or of a WIMP signature.


\subsection{Spin-dependent coupling}

\begin{figure}
\begin{center}
\begin{tabular}{cc}
\includegraphics[%
  width=0.52\linewidth,
  keepaspectratio]{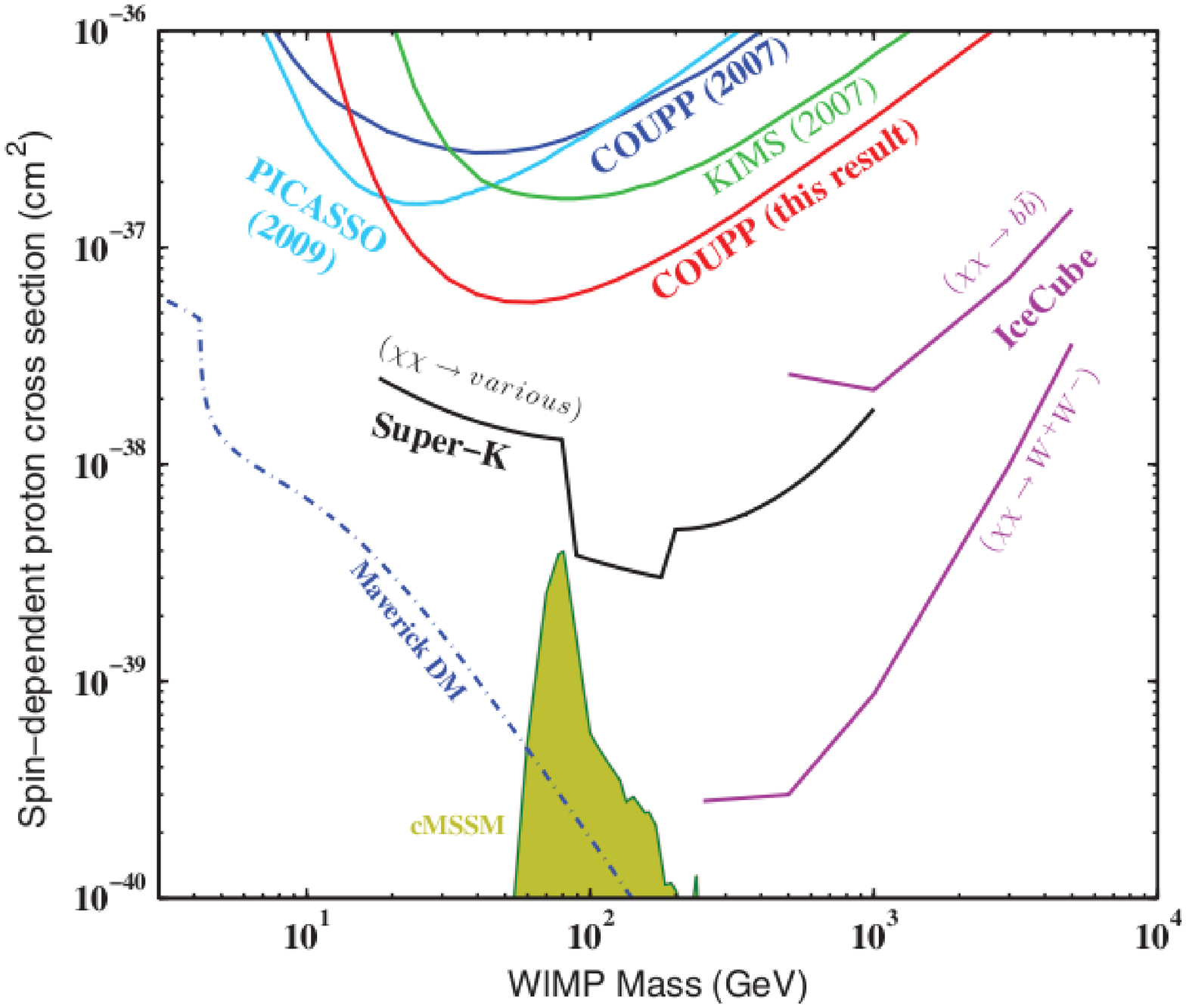}&
\includegraphics[%
  width=0.55\linewidth,
  keepaspectratio]{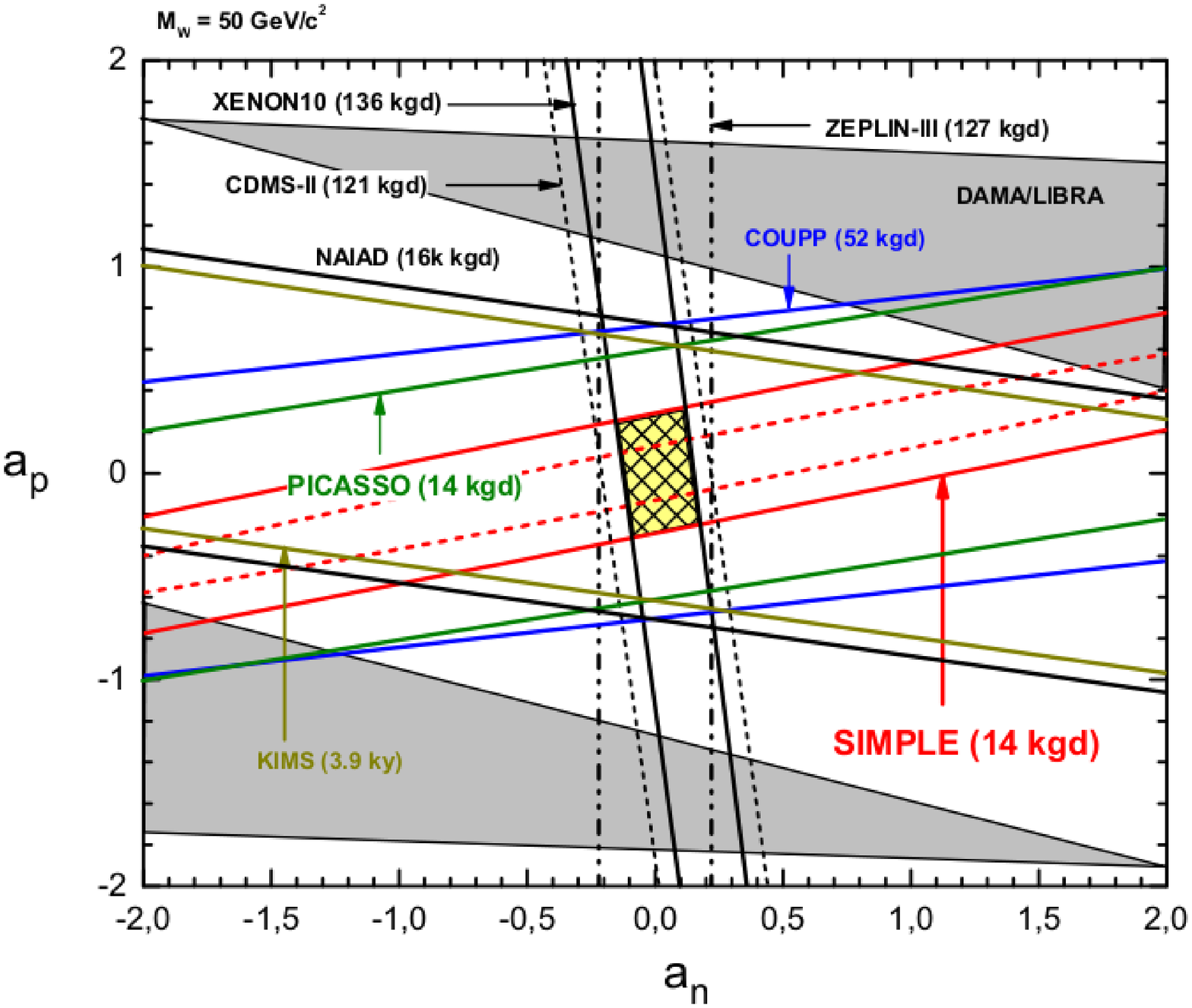}
\end{tabular}
\end{center}
\caption{Left panel: current best constraints on spin-dependent proton coupling in the MSSM phase space \cite{COUPP}. Right panel: Constraint in the proton/neutron coupling plan for several experiments \cite{PICASSO}. Latest results from COUPP not included.}
\label{spin_dependent_exclusions}
\end{figure}

Spin-dependent interactions rate is related to the nuclear spin of target nuclei.
Concerning neutron coupling, the best constraints are set by experiments that are also competitive in the spin-dependent search (like e.g. XENON or CDMS, see figure \ref{spin_dependent_exclusions}, right panel). Concerning proton coupling, direct detection methods are still far from constraining the most optimistic projections of MSSM models (see figure \ref{spin_dependent_exclusions}, left panel). Indirect methods still put the best constraints, like Kamiokande study of solar neutrino production. There are nevertheless some interesting direct detection methods that could have the ability or reaching a comparable or better level of sensitivity. There are indeed at least four different technologies ahead for continuing the search: scintillation (KIMS), bubble chamber (COUPP) and super heated droplets (PICASSO). These two lasts are using metastable states for detection. We will here focus on COUPP, showing encouraging results and giving a nice example of features allowed by such metastable detectors, not pictured in the general scheme given in figure \ref{detection_channels}.
COUPP is using one of the oldest method from accelerator physics, namely bubble chamber. Latest results have been obtained on a 3.5 kg mass of liquid CF$_{3}$I kept in a metastable state by controlling pressure. A nuclear recoil type energy deposit will locally create a bubble that can expand until full boiling of the vessel on a 100 ms time scale. One of the most impressive features of this method is its built-in capacity to reject ionising background. The bubble nucleation process is indeed almost unsensitive to electron recoils, allowing to reach a record immunity to ionising background of the order of 10$^{-10}$ (one ionising event over 10$^{-10}$ may not be rejected, where as best other experiments are at the 10$^{-5}$ level). The setup is operated at room temperature. The trigger and event positioning features are realized through the use of 2 CCD cameras monitoring the appearance of bubbles in the vessel. A particular care has to be taken with respect to the surface quality of the containing vessel, irregularities on this surface being a source of background nucleation rate. The event monitoring is completed by acoustic piezo-electric transducers. The most currently limiting background is coming from alpha emission, some possibilities of using ultrasonic emission structures to discriminate this being under study by the PICASSO\cite{PICASSO} and COUPP experiments. Neutron background is claimed to be efficiently rejected by monitoring multiple deposits (several bubbles seen on CCDs), given the high multiple interactions probability of an incoming neutron.
Latest results\cite{COUPP} gave the best constraints on proton coupling for direct methods, thanks to a 4 months run that has resulted in a final 28.1 kg.days exposure (42$\%$ quality cuts due to intermittent noise on the acoustic channel). 3 nuclear recoils candidates have been observed during this run, and a next phase with a total 60 kg bubble chamber is expected to start in 2012.

\subsection{The low-mass hints}

Back to a decade ago, the claim from DAMA of detecting an annual modulation of events seens in there NaI scintillation detectors have been widely ctiticized, their results being incompatible with results from other experiments, at least in the "standard" MSSM framework. Since then, this modulation has been confirmed by the DAMA/LIBRA experiment on several years with a correct phase and a high statistical significance\cite{DAMA}. This of course rose the question of possible long term systematic effects linked to seasonnal variations of some environmental parameter. The latter concern is all the more important that the detection has been made in the very first energy bins above threshold, the stability of this threshold being one of the most obvious questions to be adressed.
The latest results from the CoGeNT experiment\cite{CoGeNT}, claiming an excess event rate at low energy with respect to a background model and also a possible annual modulation of this excess (a claim recently fragilised by the declaration of a bias due to data cuts at the TAUP2011 conference), triggered an additional interest. The claimed modulations fall in the same region of the MSSM phase space corresponding to so-called "low-mass" candidates with a mass of the order of 10 GeV, well below previous constraints set by accelerator physics.
As already stated, one could note that both experiments are not using any discrimination of ionising background, the possibility of explaining the modulation in terms of modulated background is thus opened.
Nevertheless, a still more recent result from CRESST\cite{CRESST} gave a new intriguing hint about that. This experiment does use ionising background discrimination thanks to the simultaneous measurement of heat and scintillation in C$_{a}$WO$_{4}$ absorbers. The presence of 3 different target nuclei, together with some careful calibrations of their respective light yiels, allow to get some additional statistical informations on nuclear recoils. The three nuclear recoils have indeed different characteristic scintillation-over-heat ratio for nuclear recoils, and are sensitive to different ranges of WIMP masses.
Their latest results show a significant detection of about 30 nuclear recoils events that are also compatible with a low mass WIMP candidate. Some questions stays open about a possible contaminations from alphas events, although the background estimations made so far are well below the observed rate.
Despite of these converging results, a strong tension still exist with other experiments, XENON\cite{XENON} claiming to reject the whole MSSM phase-space zone correponding to the detected candidates (let apart a controversy about low energy calibrations in XENON), CDMS\cite{CDMS} rejecting almost all this zone too, and EDELWEISS\cite{EDELWEISS} confirming the rejection for the major part.


\begin{thebibliography}{99}



\bibitem[(Armengaud \etal\ 2011)]{EDELWEISS} Armengaud, E. \etal\ 2011, PLB, 702, 329. 
\bibitem[(Angloher \etal\ 2011)]{CRESST} Angloher, G. \etal\ 2011, arXiv:1109.0702v1.
\bibitem[(CDMS collaboration 2010)]{CDMS} The CDMS-II collaboration  2010, Nature, 327, 1619.
\bibitem[(XENON 100 collaboration 2011)]{XENON}  The XENON 100 collaboration 2011, arXiv:1104.2549v3.
\bibitem[(Aalseth \etal\ 2011)]{CoGeNT} Aalseth, CE. \etal\ 2011, arXiv:1106.0650v1.
\bibitem[(Bernabei \etal\ 2010)]{DAMA} Bernabei, R. \etal\ 2010, Eur. Phys. J. C, 67, 39.
\bibitem[(Behnke \etal\ 2011)]{COUPP} Behnke, E. \etal\ 2011, PRL 106, 021303.
\bibitem[(Archambault \etal\ 2009)]{PICASSO} Archambault, S. \etal\ 2009, PLB, 682, 185.


\end{thebibliography}
\end{document}